\documentclass[prd,aps,showpacs,preprintnumbers,amssymb]{revtex4}
\usepackage{axodraw}
\usepackage{color}
\usepackage{epsf}

\def\e3p{$\eta \rightarrow 3 \pi$}

\begin{document}
\title{%
\hfill{\normalsize\vbox{%
\hbox{}
 }}\\
{Note about the implementation of finite symmetries in the lepton sector}}
\author{Renata Jora
$^{\it \bf a}$~\footnote[1]{Email:
 rjora@theory.nipne.ro}}

\author{M. Naeem Shahid
$^{\it \bf b}$~\footnote[3]{Email:
   mnshahid@phy.syr.edu }}

\affiliation{$^{\bf \it a}$ National Institute of Physics and Nuclear Engineering PO Box MG-6, Bucharest-Magurele, Romania}

\affiliation{$^ {\bf \it b}$ National Centre for Physics,
 Quaid-i-Azam University Campus, 45320 Islamabad, Pakistan}

\date{\today}

\begin{abstract}
Assuming that the leptonic mixing matrix is given in first approximation by the tribimaximal form or a generalization of this we deduce some consequences with regard to the finite symmetries that may be applied in the minimal lepton sector of the electroweak theory.
\end{abstract}
\pacs{14.60.Pq, 12.15F, 13.10.+q}
\maketitle

\section{Introduction}

Important information about the neutrino masses and mixings  has been determined recently by a group of remarkable experiments \cite{K}-\cite{Daya}.  A large number of theoretical models \cite{King}-\cite{Jora4} have been constructed based on this experimental data and many of them considered the implementation of discrete theories in the leptons sector of the electroweak theory.

One theoretical approximation of the leptonic mixing matrix suggests that this is very close to the so-called tribimaximal form.
\begin{eqnarray}
R=
\left[
\begin{array}{ccc}
-2/\sqrt{6}&1/\sqrt{3}&0\\
1/\sqrt{6}&1/\sqrt{3}&1/\sqrt{2}\\
1/\sqrt{6}&1/\sqrt{3}&-1/\sqrt{2}
\end{array}
\right].
\label{tri6665}
\end{eqnarray}

The tribimaximal form  has increased the interest in the implementation of finite group symmetries  into the lepton sector \cite{Wolfenstein}-\cite{Jora4}. Special attention has been payed to the
the $S_3$ group because the tribimaximal matrix is just the transformation matrix which decomposes the three dimensional reducible representation into a sum of two and one dimensional irreducible pieces.

In \cite{Jora1} we applied this symmetry as a zeroth order approximation into the electroweak theory and proved that it is impossible to obtain the tribimaximal form if both neutrinos and left and right handed charged leptons are in the same three dimensional representation of the symmetry group. In the present work we will show that one can prove  similar statements for a whole class of models which employ finite groups in the lepton sector.

Consider that part of the Lagrangian which contains the lepton fields and their interaction with the $SU(2)_L\times U(1)$ gauge bosons:
\begin{eqnarray}
{\cal L}=-{\bar L}_a\gamma_{\mu}{\cal D}_{\mu}L_a-{\bar e}_{Ra}\gamma_{\mu}{\cal D}_{\mu}e_{Ra}
\label{lagr444}
\end{eqnarray}
Here,
\begin{eqnarray}
L_a=
\left[
\begin{array}{c}
\rho_a\\
e_{La}
\end{array}
\right],
\label{l665}
\end{eqnarray}

and $\rho_a$ is a two component neutrino field.

The charged leptonic weak interaction can be read from the term:
\begin{eqnarray}
{\cal L}=\frac{i g}{\sqrt{2}}W_{\mu}^{-}{\hat {\bar{e}}}_L\gamma_{\mu}K\hat{\rho}+h.c.+...,
\label{lept7776}
\end{eqnarray}

where g is the weak coupling constant, $W_{\mu}^-$ is the charged vector intermediate boson and K is the leptonic mixing matrix.

The hatted fields correspond to the mass eigenstates and can be written in terms of the gauge eigenstates as:
\begin{eqnarray}
\rho=U\hat{\rho}
\hspace{1cm}
e_L=W{\hat{e}}_L.
\label{gg55}
\end{eqnarray}

The leptonic mixing matrix K is then given by:
\begin{eqnarray}
K=W^{\dagger}U.
\label{lept66}
\end{eqnarray}

\section{Connection between representations, eigenvalues and eigenstates}

Consider an arbitrary Majorana mass term for neutrinos.
\begin{eqnarray}
{\cal L}_{mass}=-\frac{1}{2}\rho^T\sigma_2M_{\nu}\rho+h.c.+....
\label{mass444}
\end{eqnarray}

We implement a finite symmetry into the neutrino sector. Our arguments apply as well to Dirac neutrinos and charged leptons provided the left handed and right handed states transform in the same way. Since there are three neutrino states these must be in a three dimensional representation of the group.
This representation can be irreducible, reducible to a sum of a one dimensional and a two dimensional representations or reducible to a sum of three one dimensional irreducible representations. The first case leads to three degenerate eigenvalues, second to two degenerate eigenvalues and one different and the third one to three different eigenvalues.
Moreover when the representation is reducible the diagonalization matrix is just the transformation matrix between the reducible representation and the corresponding sum of irreducible representations.  There is simple proof to the above statements that we will give below.
Assume the matrix U realizes the diagonalization of the mass matrix such that:
\begin{eqnarray}
U^{\dagger}M_{\nu}U=\hat{M}_{\nu},
\label{diag7776}
\end{eqnarray}

where $\hat{M}_{\nu}$ is the diagonal matrix of eigenvalues or square of eigenvalues depending on the diagonalization procedure.

If $D(a)$ are the $3\times3$ matrices corresponding to the elements a of the finite group G in its three dimensional representation
then the invariance under the group G means that:
\begin{eqnarray}
D(a)^{\dagger}M_{\nu}D(a)=M_{\nu}.
\label{reps}
\end{eqnarray}

Then we can write:
\begin{eqnarray}
&&U^{\dagger}D(a)^{\dagger}M_{\nu}D(a)U=\hat{M}_{\nu}
\nonumber\\
&&(U^{\dagger}D(a)^{\dagger}U)\hat{M}_{\nu}(U^{\dagger}D(a)U)=\hat{M}_{\nu}.
\label{res5546}
\end{eqnarray}

The representation D(a) must be unitary as it is required by the invariance of the Lagrangian which leads to:
\begin{eqnarray}
[\hat{M}_{\nu},U^{\dagger}D(a)U]=0.
\label{res6665}
\end{eqnarray}

One can deduce from Eq.(\ref{res6665}) that if the matrix $U^{\dagger}D(a)U$ is a full $3\times3$ matrix all eigenvalues must be equal; if the matrix $U^{\dagger}D(a)U$ is a block diagonal matrix of type $(2+1)$ two eigenvalues are equal and one can be different; if the matrix $U^{\dagger}D(a)U$ is diagonal all three eigenvalues can be different. First case corresponds to a
irreducible representation, second and third to the decomposition:
\begin{eqnarray}
U^{\dagger}D(a)U={\rm block\,\,diagonal}
\label{res44443333}
\end{eqnarray}

where the block diagonal form corresponds to the two possible decompositions:
\begin{eqnarray}
&&3=2+1
\nonumber\\
&&3=1+1+1
\label{dec4443}
\end{eqnarray}

\section{Finite groups and tribimaximal leptonic mixing matrix}

The left handed states for both neutrino and charged leptons (or the neutrino state and the left handed charged lepton for a Majorana neutrino) must be in the same representation of the finite group such that to form group invariants in the Lagrangian. On the other hand the right handed states could be  in any other representation case in which the invariance under the finite symmetry is realized in conjunction with the Higgs fields. In such cases when the left handed states and the right handed states belong to different representations  the transformation matrices from the gauge eigenstates to the mass eigenstates can have any form and so does the leptonic mixing matrix. We will not discuss this case here.

If both left handed state and right handed states for each lepton species transform in the same way under the finite group G there are two possible cases:

Case I:

Both  neutrinos and charged leptons transform under the same three dimensional irreducible representation of the group. In this case the eigenvalues for both neutrinos and charged leptons are completely degenerate.
The transformation matrices are the unit matrix and the leptonic mixing matrix is also the unit matrix. Unless some perturbations are applied one cannot obtain the leptonic mixing matrix of the tribimaximal form.

Case II.

Both neutrinos and charged leptons transform under the same three dimensional reducible representation of the group. Here there are two possibilities for the decomposition of the reducible representation:
\begin{eqnarray}
&&3=2+1
\nonumber\\
&&3=1+1+1
\label{repr6665}
\end{eqnarray}

 Here the one dimensional representations may be different.

The diagonalizing matrices are just those matrices which transform from the basis of reducible representation to the basis where the representation can be written as a sum of two or three irreducible pieces.

 For the first situation in Eq. (\ref{repr6665}) there is one invariant vector corresponding to the one dimensional irreducible representation. This vector is contained both in U and in W.
 Let us write $U=[u_1,u_2,u_3]$ and $W=[w_1,w_2,w_3]$ where $u_i$ and $w_i$ are three dimensional vectors. Assume $u_1$ and $w_2$ are the invariant vectors for both cases. Then the product of $u_1$ with the vectors in W will lead in $W^{\dagger}U$ to the second row of the type $(1,0,0)$ which cannot exist in a tribimaximal matrix. More elegantly this can be seen from:
 \begin{eqnarray}
 (1+2)\times(1+2)=1\times1+1\times2+2\times1+2\times2
 \label{554}
 \end{eqnarray}

 In order to have a tribimaximal form for the leptonic mixing matrix one would need a reducible three representation in the product above.

 A similar argument applies for the second case in Eq. (\ref{repr6665}). The three one dimensional irreducible representations correspond to three invariant vectors in U an W which will lead in the product $W^{\dagger}U$ to a matrix which has three rows $(1,0,0)$, $(0,1,0)$ and $(0,0,1)$ in an arbitrary order. This can also be deduced from:
 \begin{eqnarray}
 (1+1+1)\times(1+1+1)=1\times1+1\times1+1\times1+1\times1+1\times1+1\times1+1\times1+1\times1+1\times1
 \label{res4443}
 \end{eqnarray}

We just proved a no-go theorem for the tribimaximal mixing:

  \textit{If  a finite symmetry is applied in the minimal lepton sector of the electroweak theory and if the left handed and right handed states (for the neutrinos and charged leptons separately) transform in the same representation of the group, the leptonic mixing matrix cannot have the tribimaximal form no matter which group is considered.
  }
 \section{A positive statement}

 Consider that part of the electroweak Lagrangian which contains the mass terms for the neutrino and charged leptons and also the charged current leptonic weak intercation.
 An arbitrary symmetry can be implemented into this sector.  We change from the gauge eigenstate basis to the mass eigenstate. The leptonic mixing matrix is as before
 $W^{\dagger}U$.  Assume that in first order this matrix has the tribimaximal from $K\approx R$. Then we make another unitary change of basis:
 \begin{eqnarray}
 &&\rho'=K\hat{\rho}
 \nonumber\\
 &&e_L'=e_L
 \label{ch6665}
 \end{eqnarray}

 Then the terms which contain the covariant derivatives are still invariant as the transformation is unitary, the mass term for the charged leptons will remain diagonalized and the mass term for the neutrinos will become:
 \begin{eqnarray}
{\cal L}_{mass}=-\frac{1}{2}\rho^{'T}\sigma_2K\hat{M}_{\nu}K^T\rho'+h.c.+...
\label{new4443}
\end{eqnarray}

Here $\hat{M}_{\nu}$ is the diagonal not necessarily real eigenvalue matrix. assuming that the three eigenvalues are $m_1$, $m_2$ and $m_3$ the new mass matrix will have the form:
\begin{eqnarray}
K\hat{M}_{\nu}K^T=
\left[
\begin{array}{ccc}
2/3m_1+1/3m_2&-1/3m_1+1/3m_2&-1/3m_1+1/3m_2\\
-1/3m_1+1/3m_2&1/6m_1+1/3m_2+1/2m_3&1/6m_1+1/3m_2-1/2m_3\\
-1/3m_1+1/3m_2&1/6m_1+1/3m_2-1/2m_3&1/6m_1+1/3m_2+1/2m_3\\
\end{array}
\right].
\label{f5554}
\end{eqnarray}

 If $m_1=m_3$  in $\hat{M}_{\nu}$ then the matrix $K\hat{M}_{\nu}K^T$ is the most general mass matrix invariant under the $S_3$ group. In all other cases the matrix $K\hat{M}_{\nu}K^T$ is invariant under an $S_2$ group.
One can consider also the left handed charged  leptons invariant under one of these groups, in conjunction or not with the Higgs fields.

In any model of the lepton masses and mixings for which as a first approximation the leptonic mixing matrix has the tribimaximal form there is an accidental or effective $S_3$  (if the first and third eigenvalues are degenerate) or $S_2$ symmetry (in all other cases) which may or not be broken explicitly by the right handed charged leptons. Thus finite groups which are isomorphic to or contain $S_2$ or $S_3$ as subgroups are one of the best candidates to describe this sector of the electroweak theory under this assumption.

\section*{Acknowledgments} \vskip -.5cm

Work of R. J. was supported by PN 09370102/2009.

\end{document}